\documentclass[referee]{raa}            % referee version: for submission

\usepackage{graphicx,times}  
\usepackage{natbib}           
\usepackage[a4paper, total={6in, 8in}, top=1.28in, bottom=0.6in]{geometry}

\def\co{$^{13}$CO }

\def\hi{\textsc{Hi} }
\def\hino{\textsc{Hi}}

\newcommand{\kms}{km s$^{-1}$ }
\newcommand{\kmsno}{km s$^{-1}$}
\newcommand{\as}{$^{\prime\prime} $}
\newcommand{\am}{$^{\prime} $}
\newcommand{\asno}{$^{\prime\prime}$}

\begin{document}

   \title{The revised distance of supernova remnant G15.4+0.1
%\footnotetext{$*$ Supported by the National Natural Science Foundation of China.}
}
%   \subtitle{I. Place Your Subtitle Here}

   \volnopage{Vol.0 (200x) No.0, 000--000}      %%preserved for Editor. DOn't remove!
   \setcounter{page}{1}          %%starting page, preserved for Editor. DOn't remove!

   \author{H. Su \inst{1, 2, 3},   
          M. F. Zhang \inst{1, 2},
          H. Zhu \inst{1},
          \and
          D. Wu \inst{1, 4}
   }

   \institute{Key Laboratory of Optical Astronomy, National Astronomical Observatories, Chinese Academy of Sciences, Beijing 100012, China; {\it hq\_su@bao.ac.cn, zhuhui@bao.ac.cn }\\
        \and
             University of Chinese Academy of Science, 19A Yuquan Road, Beijing 100049, China\\
        \and
             International Centre for Radio Astronomy Research, Curtin University, Bentley, WA 6102, Australia\\
        \and
             College of Information Science \& Technology, Beijing Normal University, Beijing 100875, China
   }

   \date{Received~~2017 May 26; accepted~~2017~~month day}

\abstract{
% aims
We measure the distance to the supernova remnant G15.4+0.1 which is likely associated with TeV source HESS J1818-154.
% methods
We build the neutral hydrogen (\hino) absorption and \co spectra to supernova remnant G15.4+0.1  by employing the data from the Southern Galactic Plane Survey (SGPS) and the \hino/OH/Recombination line survey (THOR).
% results
The maximum absorption velocity of about 140~\kms constrains its lower limit distance of about 8.0~kpc. Further, the fact that the \hi emission feature at about 95~\kms seems no corresponding absorption suggests that G15.4+0.1 has likely an upper limit distance of about 10.5~kpc. \co spectrum to the remnant supports our measurement. The new distance revises parameters on its associated pulsar wind nebula and TeV source.
}

   \authorrunning{H. Su, M. F. Zhang, H. Zhu, \& D. Wu}            %author_head in even pages
   \titlerunning{The distance of supernova remnant G15.4+0.1}  % title_head in odd pages

   \maketitle
%% The author head (on even pages) and the title head (on odd pages) will be
%% automatically extracted from \author{} and \title{}. Whenever the title is too long,
%% you will be asked to supply a shorter one by inserting either \authorrunning{} or
%% \titlerunning{} before \maketitle. Anyway, you can specify your own heads.
%%
%%
%% Note: In the following text body of your manuscript, please note several differences from
%%       other major journals:
%% (1) \subsection{Please Capitalize the First Letter of Each Notional Word in Subsection Title}
%% (2) Please Capitalize the First Letter of Each Notional Word in all tables' captions

%
%________________________________________________ sections below
%

\section{Introduction}
 The supernova remnant (SNR) G15.4+0.2, also called G15.42+0.18 (G15 hereafter), has a break-out morphology toward its south and a northern shell with a radius of about 6\am\ in 1420~MHz. The spectral index of G15 is $-$0.62 $\pm$ 0.03 from 330 to 4800~MHz \citep{Supan2015A&A...576A..81S, Sun2011A&A...536A..83S} with its peak flux density of 5.6~Jy at 1~GHz. G15 has a morphological correspondence with the TeV $\gamma$-ray source HESS J1818-154 \citep{Adrian-Martinez2011arXiv1112.0478A}. The high energy radiation possibly originates from a pulsar wind nebula (PWN) although the pulsar has not been detected. Multiwavelength studies have been performed to explore their properties (e.g. \citealt{Abramowski2014A&A...562A..40H, Castelletti2013A&A...557L..15C}) 

Distance is a key to further study this TeV SNR. A near distance for the remnant G15 was suggested by \citet{Castelletti2013A&A...557L..15C} by analyzing a noisy \hi absorption spectrum to the remnant. We make a further effort to obtain its reliable distance by taking advantage of new-released \hino/OH/Recombination line survey (THOR, \citealt{Beuther2016A&A...595A..32B}), the archive Southern Galactic Plane Survey (SGPS, \citealt{McClure-Griffiths2005ApJS..158..178M}), the Very Large Array Galactic Plane Survey (VGPS, \citealt{Stil2006AJ....132.1158S}), and the Galactic Ring Survey (GRS, \citealt{Jackson2006ApJS..163..145J}) data.

The SGPS has a resolution of 100\as\ and a sensitivity of less than 1~mJy~beam$^{-1}$ and were observed by The Australia Telescope Compact Array (ATCA) and Parkes telescopes. The SGPS survey has a reliable absolute flux scale by including the Parkes data. The continuum data of SGPS near G15 region is currently not available. The THOR survey observed \hino, four OH, 19 H$_{\alpha}$ recombination lines, and the L-band continuum covering the northern Galactic plane (15~$<~l~<$~67$^{\circ}$ and $|b|$~$<$~1$^{\circ}$). The THOR data has an angular resolution about 20\asno. We convert the unit of Jy~beam$^{-1}$ in the THOR data to Kelvin (K) using factors of 1536.20 for the cube data. The absolute scale of the surface brightness in the THOR data is not reliable due to lacking zero baseline. However, the relative scale is good enough for us to do \hi absorption analysis. In addition, we adopt the $^{13}$CO data from the GRS to confirm the reality of the absorption peaks, because \co is a tracer of cold and dense H$_2$ molecular cloud.

\section{Spectra analysis}
\label{sec:spec}
We use well-tested methods in order to build a reliable \hi absorption spectrum to G15 as possible \citep{Tian2007A&A...474..541T, Tian2011ApJ...729L..15T, Zhu2013ApJ...775...95Z}. 

\begin{figure}
\includegraphics[width=0.8\textwidth]{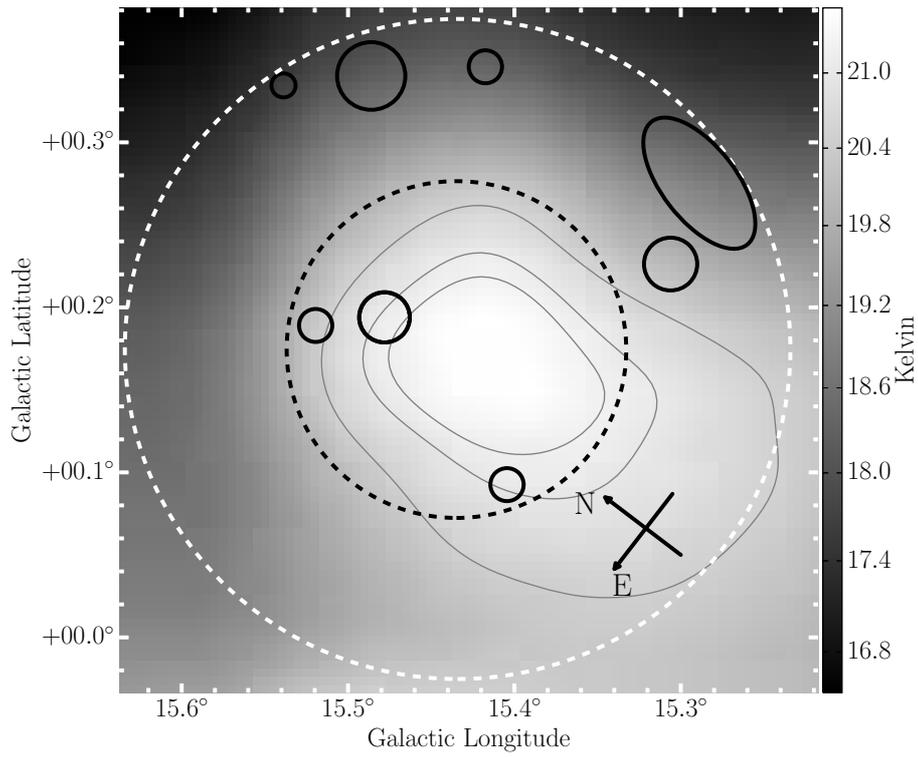}
\caption{The image of G15 at 1420~MHz from the VGPS data. The black dashed circle shows the source region (`ON' direction) with its center at (g, l) = (15.43$^\circ$, 0.17$^\circ$) and a radius of 6\farcm1. The region between the white solid circle and the white dashed circle is the background region (`OFF' direction). The black circles show point-like sources from the THOR continuum image, which are excluded from our analysis. The black solid ellipse shows a bright source in the SGPS \hi channel maps, which is also flagged in our spectra analysis. The gray contour levels are 20.5, 21.0, and 21.2 Kelvin.}
\label{fig:g15_img} 
\end{figure}

\begin{figure}
\includegraphics[width=0.8\textwidth]{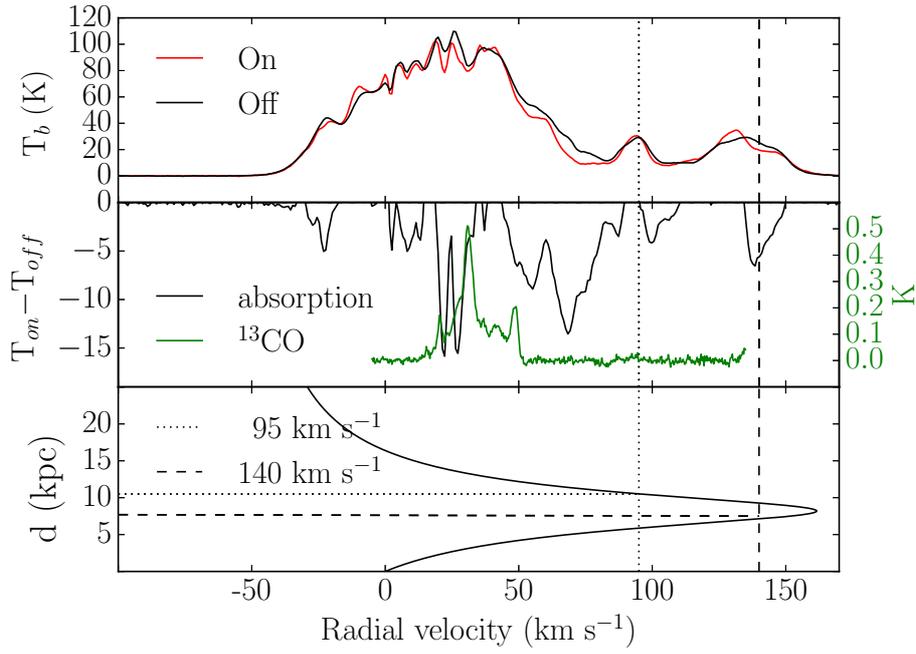}
\caption{The \hi emission, \hi absorption, and the $^{13}$CO spectra of G15 from SGPS and GRS data. The bottom panel shows the relationship between the radial velocity and the distance.}
\label{fig:sgps} 
\end{figure}

\begin{figure}
\includegraphics[width=0.8\textwidth]{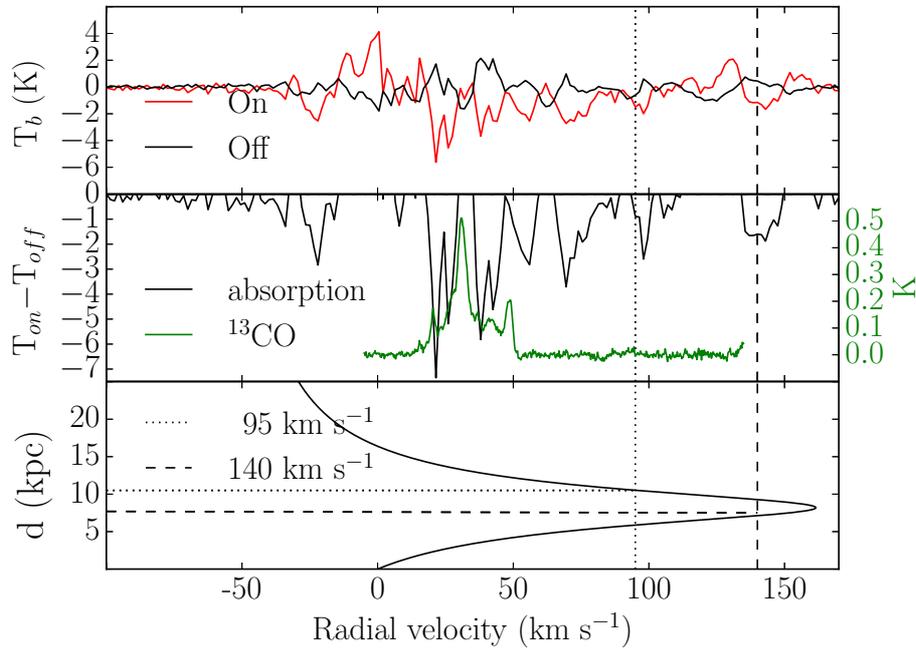}
\caption{The \hi emission, \hi absorption, and the $^{13}$CO spectra of G15 from THOR and GRS data.}
\label{fig:thor} 
\end{figure}

We improve the absorption spectrum extraction by two steps. We choose a background region which is an annulus around the source region. Our background is adjacent to the source region, which makes that the background spectrum in the source direction. We remove six point-like sources from both the continuum map and the data cube to avoid contamination (see Fig.~\ref{fig:g15_img}). We show the \hi emission, the \hi absorption, and the \co emission spectra of G15 (see Fig.~\ref{fig:sgps} and~\ref{fig:thor}). We extract the \hi emission spectra from our source and background regions and then calculate the difference between the two spectra. This difference spectrum is actually proportional to the absorption spectrum, which is used to do absorption analysis. We extract the \co spectrum in our selected background region from the GRS data and compare with our absorption spectrum.

The maximum velocity \hi absorption feature appears at about 140~\kms (see Fig.~\ref{fig:sgps} and~\ref{fig:thor}). The \co has data between -5 \kms and 135 \kmsno. There is, in fact, a \co emission feature at 135 \kms which seems corresponding with the \hi maximum velocity absorption feature. The rms of 0.009~K is obtained by averaging no \co emission-line velocity ranges from -5 to 10~\kms and from 60 to 120~\kmsno. The \co peak temperature at 135~\kms is 0.047~K, which is a 5$\sigma$ detection. This supports the \hi absorption feature at about 140~\kms is real. The velocity of 140~\kms corresponds to a distance of about 8~kpc based on the Galactic rotation model with $\Theta_{\odot}$ = 220~\kms and R$_{\odot}$ = 8.5~kpc \citep{Fich1989ApJ...342..272F}. Thus, the lower limit distance of G15 is 8~kpc.

The \hi emission at a velocity of about 95~\kms seems no responsible absorption, which means G15 is likely located in front of this \hi cloud. This velocity corresponds to a far side distance of about 10.5~kpc. The absorption peak at the velocity of about -20~\kms is possibly caused by \hi emission fluctuation so likely not a real feature (see the \hi emission spectra in Figs.~\ref{fig:sgps} and~\ref{fig:thor}). Similar fake absorption feature is often seen in faint SNRs (e.g. \citealt{Leahy2008A&A...480L..25L}).

\section{Results and brief discussion}
We obtain a new distance of 9.3~$\pm$~1.3~kpc for G15. This is consistent with the 10 $\pm$ 3~kpc derived from the surface brightness vs. diameter relation of Galactic SNRs \citep{Adrian-Martinez2011arXiv1112.0478A}. 
\citet{Castelletti2013A&A...557L..15C} showed an \hi absorption spectrum and suggested a smaller distance for the remnant. However, we do not find a clue from their paper to reproduce their spectrum, so we believe our careful measurement is more reliable. 

The \hi shell of G15 is 16.5~pc in radius (for 6\farcm1 arcmin) at a distance of 9.3~kpc. The PWN inside G15 has a total energy of less than 1.2 $\times$ 10$^{50}$~erg scaled from the value derived in \citet{ Abramowski2014A&A...562A..40H}.

We revise the parameters of G15 using the shell-type model of SNR in \citet{Sturner1997ApJ...490..619S}. We assume the progenitor supernova of G15 has an explosion energy of 10$^{51}$~erg, a mass of 8~M$_{\odot}$, a magnetic field strength of 1~$\mu$G, and an interstellar medium (ISM) density of 7 hydrogen atoms per cube centimeter, so we get the expansion velocity of about 700~\kmsno, an age of 11~kyr, and a total atomic gas mass that
forms the G15 shell of 3.3~$\times$~10$^3$~M$_{\odot}$. The estimated parameters may have a 50 percent uncertainty due to the large variance of the interstellar medium density. 

\section*{Acknowledgements}
We thank Dr. Wenwu Tian for his initial questioning the SNR G15.4+0.1's distance issue and helpful comments during our writing the paper. We acknowledge support from the NSFC (11473038, 11273025). This publication makes use of molecular line data from the Boston University-FCRAO Galactic Ring Survey (GRS).

\label{lastpage}

\bibliographystyle{raa}
\bibliography{/home/aquan/Dropbox/references/bibtex}

\end{document}